\let\new=\newcommand
\new{\eq}{\begin{equation}}
\new{\en}{\end{equation}}
\new{\degree}{$^\circ$ }

\documentclass[iop]{emulateapj}

\shorttitle{Stellar Companion to Exoplanet Host HD~177830}
\shortauthors{Roberts et al.}

\begin{document}

%\received{ } 
%\accepted{ } 

\title{Know the Star, Know the Planet. V. Characterization of The Stellar Companion to the Exoplanet Host Star HD~177830}
  
\author{
Lewis C. Roberts Jr.\altaffilmark{1},
Rebecca Oppenheimer\altaffilmark{2}, 
Justin R. Crepp\altaffilmark{3}, 
Christoph Baranec\altaffilmark{4},
Charles Beichman\altaffilmark{1,5,6}, 
Douglas Brenner\altaffilmark{2}, 
Rick Burruss\altaffilmark{1},
Eric Cady\altaffilmark{1}, 
Statia Luszcz-Cook\altaffilmark{2}, 
Richard Dekany\altaffilmark{5}, 
Lynne Hillenbrand\altaffilmark{5},
Sasha Hinkley\altaffilmark{7},
David King\altaffilmark{8}, 
Thomas G. Lockhart\altaffilmark{1}, 
Ricky Nilsson\altaffilmark{2,9}, 
Ian R. Parry\altaffilmark{8}, 
Laurent Pueyo\altaffilmark{10,11}, 
Anand Sivaramakrishnan\altaffilmark{9}, 
R\'emi Soummer\altaffilmark{9},
Emily L. Rice\altaffilmark{12}, 
Aaron Veicht\altaffilmark{2}, 
Gautam Vasisht\altaffilmark{1}, 
Chengxing Zhai\altaffilmark{1},
Neil T. Zimmerman\altaffilmark{13}
}

\altaffiltext{1}{Jet Propulsion Laboratory, California Institute of Technology, 4800 Oak Grove Drive, Pasadena CA 91109, USA}
\altaffiltext{2}{American Museum of Natural History, Central Park West at 79th Street, New York, NY 10024, USA}
\altaffiltext{3}{Department of Physics, University of Notre Dame, 225 Nieuwland Science Hall, Notre Dame, IN, 46556, USA}
\altaffiltext{4}{Institute for Astronomy, University of Hawai\textquoteleft i at M\={a}noa, Hilo, HI 96720-2700, USA}
\altaffiltext{5}{Division of Physics, Mathematics, and Astronomy, California Institute of Technology, Pasadena, CA 91125, USA}
\altaffiltext{6}{NASA Exoplanet Science Institute, 770 S. Wilson Avenue, Pasadena, CA 911225, USA}
\altaffiltext{7}{School of Physics, University of Exeter, Stocker Road, Exeter, EX4 4QL, UK}
\altaffiltext{8}{Institute of Astronomy, University of Cambridge, Madingley Rd., Cambridge, CB3 OHA, UK}
\altaffiltext{9}{Department of Astronomy, Stockholm University, AlbaNova University Center, Roslagstullsbacken 21, SE-10691 Stockholm, Sweden}
\altaffiltext{10}{Space Telescope Science Institute, 3700 San Martin Drive, Baltimore, MD 21218}
\altaffiltext{11}{Johns Hopkins University, 3400 N. Charles Street, Baltimore, MD 21218, USA}
\altaffiltext{12}{Department of Engineering Science and Physics, College of Staten Island, City University of New York, Staten Island, NY 10314, USA}
\altaffiltext{13}{Princeton University, MAE, D207 Engineering Quad, Princeton, NJ 08544}

\email{lewis.c.roberts@jpl.nasa.gov}
 
%%%%%%%%%%%%%%%%%%%%%%%%%%%%%%%%%%%%%%%%%%%%%%%%%%%%%%%%%%%%%%%

\begin{abstract}

HD~177830 is an evolved K0IV star with two known exoplanets.  In addition to the planetary companions it has a late-type stellar companion discovered with adaptive optics imagery.  We observed the binary star system with the PHARO near-IR camera and the Project 1640 coronagraph.  Using the  Project 1640 coronagraph and integral field spectrograph we extracted a spectrum of the stellar companion.  This allowed us to determine that the spectral type of the stellar companion is a M4$\pm$1V.  We used both instruments to measure the astrometry of the binary system.  Combining these data with published data, we determined that the binary star has a likely period of approximately 800 years with a semi-major axis of 100-200 AU.  This implies that the stellar companion has had little or no impact on the dynamics of the exoplanets.  The astrometry of the system should continue to be  monitored, but due to the slow nature of the system, observations can be made once every 5-10 years.  

\end{abstract}

  
%%%%%%%%%%%%%%%%%%%%%%%%%%%%%%%%%%%%%%%%%%%%%%%%%%%%%%%%%%%%%%%

\section{INTRODUCTION}

HD~177830 (HIP 93746 = WDS 19053+2555) is an evolved K0IV star \citep{vogt2000} that is on the boundary between giants and subgiants \citep{ghezzi2010}.  It is a nearby star with a distance of 59.0$\pm$2.2s pc \citep{vanLeeuwen2007}. There have been several age determinations including an age of  3.8$^{+2.8}_{-1.6}$ Gyr \citep{valenti2005}, 4.03 Gyr \citep{saffe2005}, 3.28$^{+0.36}_{-0.24}$ Gyr \citep{takeda2007}, and 3.46$\pm$0.29 Gyr  \citep{jofre2015} all of which are consistent.   \citet{baines2008a} measured the diameter of the star interferometrically with the CHARA Array as 2.99$\pm$0.39 R$_\sun$ and estimated the T$_{eff}$ as 4804 K.  Using high resolutions visible spectra, \citet{mortier2013} measured the metallicity, log $g$, T$_{eff}$, $\xi_t$ and then used these parameters to estimate the mass using theoretical isochrones and a Bayesian estimation method \citep{desilva2006}.  This resulted in a mass of 1.17$\pm$0.10 M$_\sun$, and T$_{eff}$ as 4752$\pm$79 K.  \citet{jofre2015}  carried out a similar analysis using different input spectra and modeling tools to derive the input parameters, but used the same software   to compute masses. Their estimates of 1.37$\pm$0.04 M$_\sun$ and  T$_{eff}$ of 5058$\pm$35 K are slightly different. The system has been observed with multiple instruments on the Spitzer telescope and none of the observations have turned up any evidence of a debris disk \citep{beichman2005, trilling2008, bryden2009, tanner2009, dodson-robinson2011}. 

By monitoring radial velocity (RV) variations, \citet{vogt2000} discovered a planet orbiting HD~177830  with a period of 391.6 days, an eccentricity of 0.41 and an $M\sin i$ of 1.22M$_{J}$, where $M$ is the mass and $i$ is the inclination.   This planet was designated HD~177830b. It orbits the primary at a distance of 0.63-1.57 AU, and lies inside the habitable zone.  The planetary orbit was updated by \citet{butler2006} and again by \citet{wright2007} who also showed the first signs of a second planet in the system.    The star continued to be observed and \citet{meschiari2011} published an update of the  orbital parameters of HD~177830b and announced the discovery of an inner planet in the system.  This planet, HD~177830c, has a period of 110.9$\pm$0.1 days and eccentricity of 0.3$\pm$0.1 and an $M\sin i$ of 0.15$\pm$0.02 M$_J$.  The updated parameters of HD~177830b are a period of 407.31 days, an eccentricity of 0.009$\pm$0.004 and an $M\sin i$ of 1.48M$_{J}$.  \citet{reffert2011} were able to use the lack of astrometric detection with Hipparcos satellite to place an upper limit on the mass of HD~177830b of 225.2 M$_{J}$. 
 
\citet{eggenberger2007} first reported the discovery of a stellar companion to HD~177830 with a separation of 1\farcs6 using adaptive optics (AO) on the Very Large Telescope.  The stellar companion was designated HD~177830 B, while the primary was designated HD~177830 A.   Based on near-IR colors and the differential magnitude of 6.6  at 1.6 \micron~they estimated the spectral type of HD~177830 B to be M2V--M5V with a mass of 0.23$\pm$0.01 M$_\sun$.     \citet{roberts2011} published an observation of the same companion that predates the discovery images of \citet{eggenberger2007} using the visible-light AO system on the Advanced Electro-Optical System (AEOS) telescope. The astrometry from both papers is shown in Table \ref{results}. 

There have been several prior high angular resolution observations of the HD~177830. \citet{baines2008b} used the CHARA Array to search for stellar companions and did not detect the stellar companion because it was outside of the instrument's field of view of 0\farcs1. Inside of their field of view, they were able to rule out any additional stellar companions earlier than K0V.     \citet{lu1987} carried out speckle interferometry observations with the Kitt Peak 4 m telescope, but the observations did not achieve a sufficient dynamic range to detect the companion. 

The stellar companion's measured separation corresponds to a projected separation of 97AU and this separation has the potential to have an impact on the dynamical behaviour and evolution of the exoplanets orbiting HD~177830 A. The population of exoplanets in binaries with semimajor axes smaller than 100 AU is statistically different than those orbiting single stars \citep{zucker2002,bonavita2007}, while for binaries with separations larger than 100 AU the  frequencies of exoplanets among single stars and components of wide binaries  are indistinguishable \citep{raghavan2006,bonavita2007}.   With that in mind, we carried out observations with two instruments on the Palomar 5m telescope in an effort to understand the orbital parameters of this exoplanet hosting binary star.  

\section{OBSERVATIONS AND DATA ANALYSIS}\label{observations}

\subsection{Project 1640}
HD~177830 was observed with the Project 1640 (P1640) coronagraph \citep{hinkley2011} on 2012 June 12 UT.   The P1640 coronagraph is integrated with an integral field spectrograph  covering the \textit{YJH} bands.  Like PHARO, this instrument is mounted on the PALM-3000 AO system.   We collected 3 data cubes, each with an integration time of  549.9 s. For each data cube the primary was placed behind the occulting disk.  The data were reduced using the  Project 1640 data reduction pipeline \citep{zimmerman2011}.  Since that paper was written, a few upgrades have been made to the pipeline that reduce lenslet-lenslet cross talk.  The pipeline processing produces an image of the object in each of 32 wave bands resulting in a data cube.  

\begin{figure}[thb]
    \centering
    \includegraphics[height=6cm]{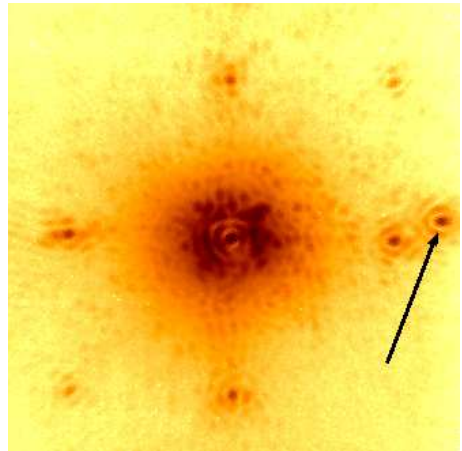} 
    \caption{A slice of the occulted P1640 image cube at a wavelength of 1.495$\micron$.  It has been rotated so that North is up and East is to the right. There are four astrometric grid spots in the image, see the text for a discussion of their use.  An arrow points towards the companion. The field of view is 4\farcs0.}%
    \label{p1640_image}%
\end{figure}

While we collected several unocculted images of HD~177830, the dynamic range between the companion and primary makes it extremely difficult to detect the companion.  Two of our occulted images used the astrometric grid spots \citep{sivaramakrishnan2006}.  We fit a Gaussian to each of these spots. Then we identified the location of the primary by fitting lines to the vertical and the horizontal spots.  The primary is located where these two lines intersect.    Figure \ref{p1640_image} shows an image slice with the grid spots.  The final astrometry is the average of the astrometry measured from 26 of the data slices from each of the two data cubes with grid spots. The frames in the water absorbtion bands with center wavelengths of 1145 nm, 1170 nm, 1395 nm, 1420 nm, 1445 nm and 1795 nm had too low signal to fit a Gaussian to the grid spots.   The astrometry is presented in Table \ref{results}, which lists the Besselian date of the observation, the position angle ($\theta$) and separation ($\rho$) of the system and the instrument used in the observation. 

\subsection{PHARO }
 
\begin{figure}[thb]
    \centering
    \includegraphics[height=6cm]{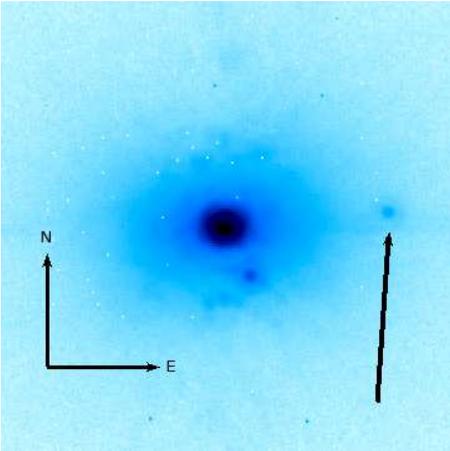}   
    \caption{Ks images of HD~177830 from the PHARO instrument taken in 2014.  An arrow points to the companion.  In the image, North is up and East is to the right.    The close object to the south east of the primary is a ghost from a neutral density filter in the camera. This is an approximately 4\arcsec~wide subimage from the full 25\arcsec field of view. } 
    \label{pharo_image}%
\end{figure}

We observed HD~177830 on 2012 May 9 UT and on 2014 May 14 UT with the Palomar Observatory Hale 5 m telescope using the PALM-3000 AO system  and the PHARO near-IR camera.    The PALM-3000 AO system is a natural guide star system using two deformable mirrors (DM) \citep{dekany2013}. One DM corrects low-amplitude high spatial frequency aberrations, while the other corrects the higher-amplitude low spatial frequency aberrations. The system is optimized for high contrast observations and routinely produces Strehl ratios  greater than 80\% in the  \textit{Ks} band.   The PHARO camera uses a HgCdTe HAWAII detector for observations between 1 and 2.5 \micron~wavelength \citep{hayward2001}. The camera has multiple filters in two filter wheels and we used the \textit{Ks} filter to collect 10 frames in 2012 and 50 frames in 2014.  In 2012, only sky frames were collected for calibration purposes. Past experience with PHARO has shown that its flat fields and dark frames are very stable from run to run. We used calibration data from 2013 to calibrate the 2012 images.  Calibration frames were collected on the same night as the 2014 data. The night of 2014 May 14 UT suffered from high winds resulting in poor seeing and lower image quality than is normal for P3K.   After debiasing, flat fielding, bad pixel correction and background subtraction, the frames were coadded.  The resulting images are shown in Figure \ref{pharo_image}. The \textit{fitstars} algorithm was used to measure the astrometry and photometry of the objects \citep{tenBrummelaar1996,tenBrummelaar2000}.   Photometric error bars were assigned using the technique described in \citet{roberts2005}. Astrometric error bars were set equal to the median error bar of 26 binaries measured with P3K and PHARO \citep{roberts2015}.  The resulting measurements are in Table \ref{results}. It has the same format as the P1640 results, but we also list the differential magnitude between the two stars in the $Ks$ filter.

 %%%%%%%%%%%%%%%%%%%%%%%%%%%%%%%%%%%%%%%%%%%%%%%%%%%%
 
\begin{deluxetable*}{lcccl}
\tablewidth{0pt}
\tablecaption{Measured astrometry and photometry for HD~177830. \label{results} }
\tablehead{\colhead{Epoch} & \colhead{$\theta$ ($^{\circ}$)} &\colhead{$\rho$ (\arcsec)}& \colhead{$\Delta$ M (\textit{Ks})} & \colhead{Instrument/Reference} }
\startdata
2002.5474 &   \phn84.1$\pm$1.0\phn   &   \phn1.62$\pm$0.01\phn  & ... & Roberts et al. (2011)\\
2004.4784 &   84.85$\pm$0.21  &     1.645$\pm$0.008    &... & Eggenberger et al. (2007)\\
2005.3494 &   84.60$\pm$0.39  &     \phn1.64$\pm$0.01\phn      &...  & Eggenberger et al. (2007)\\
2012.3549 &   \phn84.3$\pm$0.2\phn  & \phn1.67$\pm$0.01\phn    & 6.5$\pm$0.3   & PHARO\\
2012.4475 &   \phn86.0$\pm$0.1\phn  & \phn1.68$\pm$0.002    & ...  & P1640\\
2014.6748 &   \phn85.3$\pm$0.2\phn  &    1.67$\pm$0.01   &   6.1$\pm$0.3 & PHARO
\enddata
\end{deluxetable*}

\section{SPECTRAL TYPE DETERMINATION OF THE COMPANION}\label{spectral_type}
  
To create a spectrum of the stellar companion, we performed aperture photometry on the images with sufficient S/N in the P1640 data cube. We used the same techniques that we have used in our previous papers (e.g. \citealt{zimmerman2010,hinkley2010,roberts2012}). This was done with the \textit{aper.pro} routine which is part of the IDL astrolib\footnote{http://idlastro.gsfc.nasa.gov} and is an adaptation of \textit{DAOphot} \citep{stetson1987}. A photometry aperture and a sky annulus were centered on the companion. The  radius of the photometry aperture was set equal to 3.5$\lambda/D$ rounded to the nearest pixel size, where $\lambda$ is the central wavelength of each image and $D$ is the aperture of the telescope. This corresponds to a radius of 0\farcs18 at the center of J band and 0\farcs24 at the center of  H band. The radius of the sky anullus was set to  0.67\arcsec; this is large enough to avoid significant portions of the central PSF while also avoiding the occultation spot. The width of the sky annulus was set to 10 pixels, or 0.19\arcsec. The background was set equal to the average intensity of the pixels in the sky annulus and was subtracted from each pixel in the photometry aperture.  The spectrum is the summed power in each slice as function of wavelength. The \textit{aper.pro} code also produces a measurement error bar based on photon statistics and the error in the background estimation. The extracted spectrum is a convolution of the object spectrum and the spectral response function (SRF). The SRF includes the instrumental response and the absorption due to the Earth's atmosphere. We computed the SRF by using an unocculted observation of HD~177830~A as a reference source. We extracted the spectrum of HD~177830~A using the aperture photometry method described above.  Then we divided the extracted spectrum  by a reference spectrum of the same spectral type.   HD~177830 A is a K0IV.  We did not have a template for this spectral type, instead we used a K0V spectra  from the Infrared Telescope Facility (IRTF) Spectral Library \citep{rayner2009}. This produces an estimate of the SRF.  The companion's measured spectrum is then divided by the SRF and yields the calibrated spectrum of the companion. The portions of the spectrum that overlap with the water bands were removed, as they had large errors. 

Since the template spectrum had the incorrect luminosity class, we checked to see how much impact the incorrect luminosity class had on the resulting spectra.  We computed the spectrum of HD~177830 B with an SRF from HD~129814 (spectral type G5V), taken two nights later at a similar air mass. The resulting spectrum for HD~177830 B is the same to within the error bars. 

 The error bars on the final reduced spectra are the combination of the errors of the measured science spectra and the error on the SRF. The SRF errors come from the combination of the errors of the measured calibration spectra and the errors on the template spectrum.  See \citet{roberts2012} for further discussion on SRFs. 

After the spectrum of the companion was extracted from the P1640 data, we compared it against the spectra in the IRTF Spectral Library \citep{cushing2005, rayner2009}. These include FGKM main sequence stars and  LT brown dwarfs. The template spectra were binned and smoothed in order to produce the equivalent spectra to having the star observed by P1640.  Each spectrum was normalized by its average.  Then each template spectrum was compared against the measured spectrum using the sum of the squares of the residual (SSR) as a metric,  

\begin{equation}\label{fit}
SSR=\sum_\lambda w_\lambda (S_\lambda - R_\lambda)^2,
\label{ssr}
\end{equation}

\noindent where,  $S_\lambda$ is the measured spectrum at a given wavelength, $R_\lambda$ is the binned reference spectrum at the same wavelength and $w_\lambda$ is the weight at the wavelength. The best fit reference spectrum was the one with the minimum value to the metric.    The weights were set equal to the inverse of the computed error at each wavelength point.

\begin{figure*}[thb]
  \centering
  \includegraphics[width=140mm]{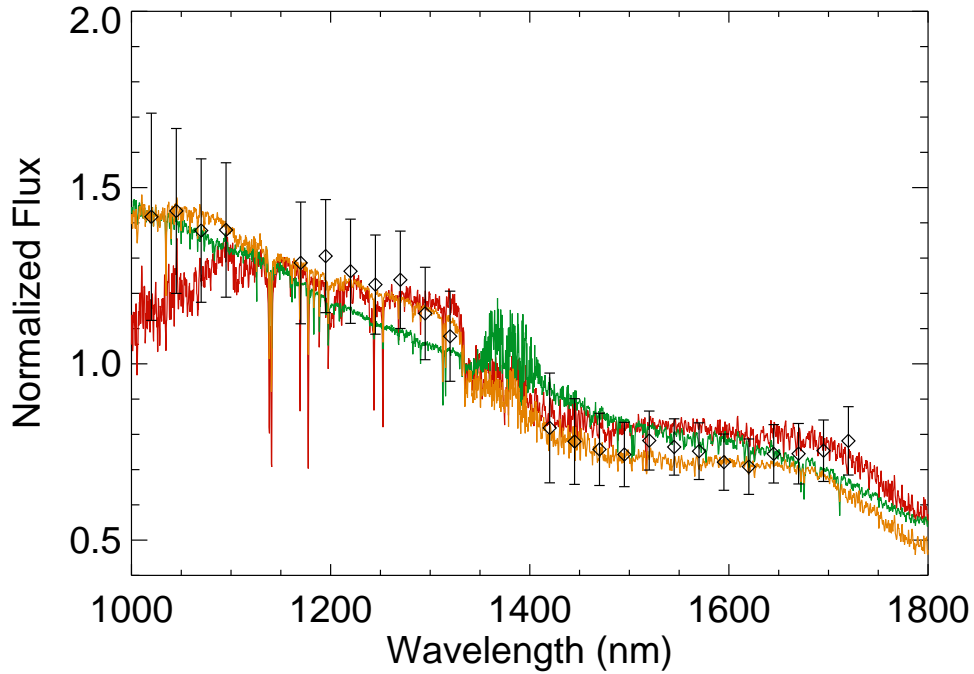}
 \caption{The spectrum extracted from the P1640 data of HD~177830 B. The three over plotted template spectra are Green=M2V, Orange=M4V,  and Red=M6V.  The determination of the spectral type is M4$\pm$1V.}
 \label{spectra}
\end{figure*}

The extracted spectrum of HD~177830~B is shown in Figure \ref{spectra}. The error bars for each data point are shown in the figure.  The error bars are smallest in the $H$-band portion of the spectrum and   since the weight of each data point is the inverse of the error bar, this means that the $H$-band data are weighted more heavily than the $J$-band data. This is appropriate for several reasons.  The AO performance is improved in the $H$-band and produces a higher quality image. The detector samples the point spread function (PSF) at a higher spatial frequency in $H$-band and finally the coronagraph is optimized for $H$-band observations.  This results in a slightly higher contrast in $H$-band \citep{hinkley2011} and lower errors.

\begin{figure}[thb]
  \centering
  \includegraphics[width=80mm]{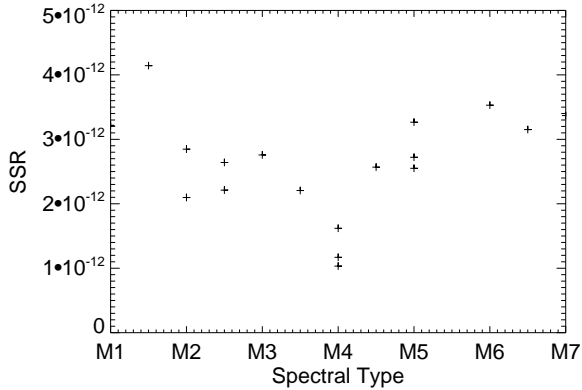}
 \caption{The SSR metric from Equation \ref{ssr} plotted as a function of spectral type for HD~177830.  The units of SSR are arbitrary.  For some spectral types, the IRTF Spectral Library has multiple stars of that class; this results in multiple data points for several of spectral types. The best fit spectral type is M4V. }
 \label{metric}
\end{figure}

In Figure \ref{spectra} we have over plotted the spectra for M2V, M4V, and M6V stars.  The best fit (via Equation \ref{ssr}) for the companion's spectral type is M4V. Figure \ref{metric} shows the SSR metric plotted as a function of spectral type.   Several adjacent spectral types have similar fitting metric values and their spectra in Figure \ref{spectra} appear to fit the data almost as well within the error bars. From this we conclude that the spectral type is M4$\pm$1V with a conservative error bar.  This agrees with the earlier photometric determination of M2V--M5V from \citet{eggenberger2007}.

\section{ORBITAL ANALYSIS}\label{orbital_analysis}

\begin{figure}[thb]
  \centering
  \includegraphics[width=80mm]{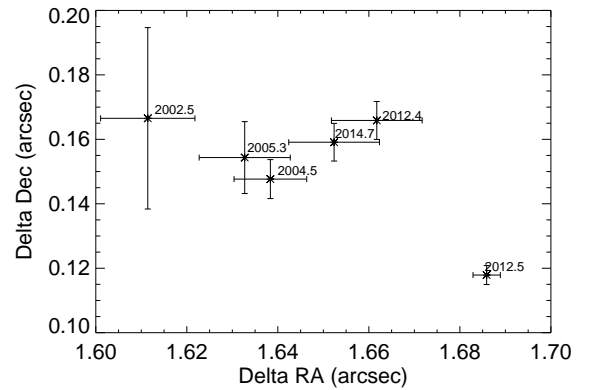}
 \caption{A plot of the relative motion of the companion to the primary star.  The primary direction of relative motion is to the East, but we can not tell if the declination is changing or not. }
 \label{motion}
\end{figure}

\citet{eggenberger2007} concluded that the binary star system had common proper motion. With the additional data from \citet{roberts2012} and  this paper, that conclusion is strengthened.  Between 2002 and 2014 the companion moved at a rate of 4.8 mas/yr relative to the primary, while if it was a fixed background star it would have been expected to move -65.92 mas/yr \citep{vanLeeuwen2007} due to the proper motion of the primary. This clearly shows that the companion has common proper motion with the primary.  We plot the relative motion in Figure \ref{motion}.  From this we can tell the  bulk of the motion is to the East, but due to the size of the error bars in relationship with  the small amount of motion, we can not tell if the companion has an increasing or decreasing declination. The 2012.4475 data point from P1640 does not line up with the rest of the data.  This is probably due to an error in the plate scale for that instrument.  The separation corresponds to approximately 90 pixels and a 1\% error in the plate scale would throw the measurement off.

Using the technique of \citet{tokovinin2014}, we can constrain the period of the binary.  Using the equation,

\eq
   P^*=\left( \frac{\rho^3 \pi^{-3}}{M_1+M_2} \right)^{1/2}
\en

\noindent where, $P^*$ is the probable period based on Kepler's third law, $\rho$ is the measured separation, assumed to be the semi-major axis, $\pi$ is the parallax, and $M_1$ and $M_2$ are the mass of each star.  \citet{tokovinin2014}   carried out simulations with random orbital phases and eccentricities and showed that the median ratio of $\rho/a$ is close to 1 depending slightly on the eccentricity distribution used in the simulations.  In all cases, the ratio never exceeds 2.  This results in the ratio of $P^*/P_{true}$ always being less than 3.17.  For the case of HD~177830, we use the latest separation measurement from Table \ref{results},  and the Hipparcos measured parallax \citep{vanLeeuwen2007}.  For masses, the spectral type of M4V that we derived for the secondary corresponds to a main sequence mass of 0.2 M$_\sun$ \citep{reid2005}. 

As discussed in the introduction, there are two literature estimates of the mass estimates from spectroscopic modeling: 1.17$\pm$0.10 M$_\sun$ \citep{mortier2013} and  1.37$\pm$0.04 M$_\sun$ \citep{jofre2015}. Since the T$_{eff}$ derived by \citet{mortier2013} matches the interferometrically derived T$_{eff}$ of \citet{baines2008a}, we used the mass estimate of \citet{mortier2013} resulting in a period estimate of 829 yrs.  The mass estimate of \citet{jofre2015} produces a period of 771 years.  

With a semi-major axis  of 100-200 AU the stellar companion will not have influenced the formation or dynamical evolution of the exoplanets unless the stellar orbit is extremely eccentric.  There is little evidence that the stellar companion has impacted the dynamics of the exoplanets, with semi-major axes of 1.2218$\pm$0.0008 AU and 0.5137$\pm$0.0003 AU.    Most of the suggested ways that binary companions interact with exoplanets result in planets with eccentric orbits \citep{kley2008} or hot Jupiters \citep{wu2003, naoz2012}. In this case, HD~177830 b has a low eccentricity of 0.009$\pm$0.004 and appears not to have had its orbit altered by the stellar companion.  The inner planet, HD~177830c,  has a low to moderate eccentricity of 0.3, but since it is inside of the orbit of HD~177830 b, it seems unlikely that the stellar companion would have effected its orbit and not that of HD~177830 b. It is possible that the system has additional undetected planets in outer orbits that have been modified by the gravitational pull of HD~177830 B.  It is also possible that the system used to have additional outer planets that were ejected  from the system due to interactions with the binary  \citep{kaib2013}.   Since the eccentricity of HD~177830b is almost zero, the system does not appear to be undergoing a Kozai cycle, implying that the relative inclination of the stellar companion's orbit and that of the planets is less than  the critical angle of 39.2\arcdeg \citep{holman1997}.   

\section{NULL RESULTS}\label{null_results}

In addition to HD~177830, we used P1640 to observe several exoplanet host stars that did not have known companions.  These observations happened during the first phase of the P1640 instrument when it was used with  PALM-3000 predecessor, PALAO \citep{dekany1997, troy2000}.  We did not detect any candidate companions in our 3\farcs4 field of view.  Table \ref{unresolved} lists the name most commonly used in reference to the exoplanet, the Hipparcos number, the Besselian date of the observations and the radius of the field of view in terms of AU.  After data reduction with the version of the LOCI speckle suppression software described in \citep{crepp2011}, all the stars had contrast curves approximately the same as that shown in that paper. The contrast is high enough to rule out additional stellar companions within the field of view. 

%%%%Fit results
\begin{deluxetable}{lrlc}
\tablewidth{0pt}
\tablecaption{ Exoplanet Host Stars with No Detected Companions\label{unresolved}}
\tablehead{ \colhead{Name}  &   \colhead{HIP}   & \colhead{Date (UT)} & \colhead{FOV Radius (AU)} }
\startdata 
HD 5319  & 4287   & 2008.8058 & 195\phn \\
HD 69830 & 40693  & 2009.2054 & 21.2\\
GJ 849   & 109358 & 2008.5242 & 15.5
%51 Peg   & 113357 & 2008.5216 & 26.5\\
\enddata
\end{deluxetable}

%%%%%%%%%%%%%%%%%%%%%%%%%%%%%%%%%%%%%%%%%%%%%%%%%%%%%%%%%%%%%%%

\section{SUMMARY}\label{summary}
   
We used the P1640 coronagraph and IFU to determine that the secondary in the exoplanet hosting binary system, HD 177830, is a M4$\pm$1V star.  We also extracted the astrometry from this measurement as well as two others taken with the PHARO instrument. Combined with published astrometry, we are able to constrain the orbit to a period of approximately 800 years and a semi-major axis of 100-200 AU.  This strongly suggests that the binary system has had little to no impact on the known exoplanets in the system. The computed orbit of the system can be improved with additional astrometry, but due to the slow motion of the system, the observations can probably be made every 5-10 years until significant orbital motion is detected.

%%%%%%%%%%%%%%%%%%%%%%%%%%%%%%%%%%%%%%%%%%%%%%%%%%%%%%%%%%%%%%%

\acknowledgements

A portion of the research in this paper was carried out at the Jet Propulsion Laboratory, California Institute of Technology, under a contract with the National Aeronautics and Space Administration (NASA). This work was partially funded through the NASA ROSES Origins of Solar Systems Grant NMO710830/102190.  Project 1640 is funded by National Science Foundation grants AST-0520822, AST-0804417, and AST-0908484.   The members of the Project 1640 team are also grateful for support from the Cordelia Corporation, Hilary and Ethel Lipsitz, the Vincent Astor Fund, Judy Vale, Andrew Goodwin, and an anonymous donor. C.B. acknowledges support from the Alfred P. Sloan Foundation. We thank the staff of the Palomar Observatory for their invaluable assistance in collecting these data. This paper is based on observations obtained at the Hale Telescope, Palomar Observatory.  This research made use of the Washington Double Star Catalogue maintained at the U.S. Naval Observatory, the SIMBAD database, operated by the CDS in Strasbourg, France and NASA's Astrophysics Data System.  
  
{\it Facilities:} \facility{Hale (PHARO), \facility{Hale (Project 1640}) }

%%%%%%%%%%%%%%%%%%%%%%%%%%%%%%%%%%%%%%%%%%%%%%%%%%%%%%%%%%%%%%%

% References


\begin{thebibliography}{}
 
\bibitem[Baines et al.(2008a)]{baines2008a}
         Baines, E.K., McAlister, H.A., ten Brummelaar, T.A., Turner, N.H., Sturmann, J., Sturmann, L., Goldfinger, P.J., \& Ridgway, S.T. 2008, \apj, 680, 728

\bibitem[Baines et al.(2008b)]{baines2008b}
         Baines, E.K., McAlister, H.A., ten Brummelaar, T.A., Turner, N.H., Sturmann, J., Sturmann, L., \& Ridgway, S.T. 2008, \apj, 682, 577

\bibitem[Beichman et al.(2005)]{beichman2005}
         Beichman, C.A., Bryden, G., Rieki, G.H. et al. 2005, \apj, 622, 1160

\bibitem[Bonavita \& Desidera(2007)]{bonavita2007}
         Bonavita, M., \& Desidera, S. 2007, \aap, 468, 721

\bibitem[Bryden et al.(2009)]{bryden2009}
         Bryden, G., Beichman, C.A., Carpenter, J.M. et al. 2009, \apj, 705, 1226
       
\bibitem[Butler et al.(2006)]{butler2006}
         Butler, R.P., Wright, J.T., Marcy, G.W., et al. 2006, \apj, 646, 505

\bibitem[Crepp et al.(2011)]{crepp2011}
         Crepp, J.R., Puyeo, L., Brenner, D., et al. 2011, \apj, 729, 132

\bibitem[Cushing, Rayner \& Vacca(2005)]{cushing2005}
         Cushing, M.C. Rayner, J.T., \& Vacca, W.D., \apj, 2005, 623, 1115
 
\bibitem[Dekany et al.(1997)]{dekany1997}
         Dekany, R., Wallace, K., Brack, G., Oppenheimer, B.R., \& Palmer, D.
1997, \procspie, 126, 269

\bibitem[Dekany et al.(2013)]{dekany2013}
         Dekany, R., Roberts, J., Burruss, R., et al. 2013, \apj, 776, 130

\bibitem[de Silva et al.(2006)]{desilva2006}
         de Silva, L., Girardi, L., Pasquini, L., et al. 2006, \aap, 458, 609

\bibitem[Dodson-Robinson et al.(2011)]{dodson-robinson2011}
         Dodson-Robinson, S.E., Beichman, C.A., Carpenter, J.M., \& Bryden, G. 2011, \aj, 141, 11

\bibitem[Eggenberger et al.(2007)]{eggenberger2007}
         Eggenberger, P., Udry, S., Chauvin, G., Beuzit, J.-L., Lagrange, A.-M., Segransan, D. \& Mayor, M. 2007, \aap 474, 273

\bibitem[Ghezzi et al.(2010)]{ghezzi2010}
         Ghezzi, L., Cunha, K., Schuler, S.C., \& Smith, V.V. 2010, \apj, 725, 721

\bibitem[Hayward et al.(2001)]{hayward2001}
         Hayward, T.L., Brandl, B., Pirger, B., et al. 2001, \pasp, 113, 105
 
\bibitem[Hinkley et al.(2010)]{hinkley2010}
         Hinkley, S., Oppenheimer, B.R., Brenner, D. et al. \apj, 712, 421

\bibitem[Hinkley et al.(2011)]{hinkley2011}
         Hinkley, S., Oppenheimer, B.R., Zimmerman, N., et al. 2011, \pasp, 123, 74

\bibitem[Holman et al.(1997)]{holman1997}
         Holman, M., Touma, J. \& Tremaine, S. 1997, Nature, 386, 254

\bibitem[Jofr\'e et al.(2015)]{jofre2015}
         Jofr\'e , E., Petrucci, R., Saffe, C., Saker, L., Artur de la Villarmois, E., Chavero, C., G\'omez, M., \& P. J. D. Mauas 2015, \aap, 574, A50

\bibitem[Kaib et al.(2013)]{kaib2013}
         Kaib, N.A., Raymon, S.N., \& Duncan, M. 2013,  Nature, 493, 381 

\bibitem[Kley \& Nelson(2008)]{kley2008}
         Kley, W., \& Nelson, R.P. 2008, \aap, 486, 617 

\bibitem[Lu et al.(1987)]{lu1987}
         Lu, P.K., Demarque, P., van Altena, W., McAlister, H., \& Hartkopf, W. 1987, \aj, 94, 1318

\bibitem[Meschiari et al.(2011)]{meschiari2011}
         Meschiari, S., Laughlin, G., Vogt, S.S., Butler, R.P., Rivera, E.J., Haghighipour, N., \& Jalowiczor, P. 2011, \apj, 727, 117

\bibitem[Mortier et al.(2013)]{mortier2013}
         Mortier, A., Santos, N. C., Sousa, S. G., Adibekyan, V. Zh., Delgado Mena, E., Tsantaki, M., Israelian, G., \& Mayor M. \aap, 557, A70

\bibitem[Naoz et al.(2012)]{naoz2012}
         Naoz, S., Farr, W.M., \& Rasio, F.A. 2012, \apjl, 754, L36

\bibitem[Raghavan et al.(2006)]{raghavan2006}
         Raghavan, D., Henry, T. J., Mason, B. D., et al. 2006, \apj, 646, 523

\bibitem[Rayner et al.(2009)]{rayner2009}
          Rayner, J.T., Cushing, M.C., \& Vacca, W.D. 2009, \apjs, 185, 289

\bibitem[Reffert \& Quirrenbach(2011)]{reffert2011}
         Reffert, S., \& Quirrenbach, A., 2011, \aap, 527, A140

\bibitem[Reid \& Hawley(2005)]{reid2005}
         Reid, I.N., \& Hawley, S.L. 2005, New Light on Dark Stars, (2nd ed; Berlin; Springer-Verlag)

\bibitem[Roberts et al.(2005)]{roberts2005}
         Roberts Jr., L.C., Turner, N.H, Bradford, L.W., et al. 2005, \aj, 130, 2262

\bibitem[Roberts et al.(2011)]{roberts2011}
         Roberts, Jr. L.C., Turner, N.H., ten Brummelaar, T.A., Mason, B.D., \& Hartkopf, W.I. 2011, \aj, 142, 175

\bibitem[Roberts et al.(2012)]{roberts2012}
         Roberts, Jr., L.C., Rice, E.L., Beichman, C.A. et al. 2012, \aj, 144, 14

\bibitem[Roberts et al.(2015)]{roberts2015}  
         Roberts, Jr., L.C., Tokovinin, A., Mason, B.D., Hartkopf, W.I., \& Riddle, R.L. 2015, \aj, Submitted

\bibitem[Saffe et al.(2005)]{saffe2005}
         Saffe, C., G\'omez, M., \& Chavero, C. 2005, \aap, 443, 609

\bibitem[Sivaramakrishnan \& Oppenheimer(2006)]{sivaramakrishnan2006}
         Sivaramakrishnan, A., \& Oppenheimer, B.R. 2006, \apj, 647, 620

\bibitem[Stetson(1987)]{stetson1987}
         Stetson, P.B. 1987, \pasp, 99, 191

\bibitem[Takeda et al.(2007)]{takeda2007}
         Takeda, G., Ford, E.B., Sills, A., Raslo, F.A., Fischer, D.A., \& Valenti, J.A. 2007, \apjs, 168, 297
 
\bibitem[Tanner et al.(2009)]{tanner2009}
         Tanner, A., Beichman, C., Bryden, G., Lisse, C., \& Lawler, S. 2009, \apj, 704, 109

\bibitem[ten Brummelaar et al.(1996)]{tenBrummelaar1996}
         ten Brummelaar T.A., Mason B.D., Bagnuolo, Jr. W.G., Hartkopf W.I., McAlister H.A.,  Turner N.H. 1996, AJ, 112, 1180

\bibitem[ten Brummelaar et al.(2000)]{tenBrummelaar2000}
         ten Brummelaar T.A., Mason McAlister H.A., Roberts Jr. L.C., Turner N.H., Hartkopf W.I.,  Bagnuolo Jr., W.G. 2000, AJ, 119, 2403

\bibitem[Tokovinin(2014)]{tokovinin2014}
         Tokovinin, A. 2014, \aj, 147, 86 

\bibitem[Trilling et al.(2008)]{trilling2008}
         Trilling, D., Bryden, G., Beichman, C.A. et al. 2008, \apj, 674, 1086
 
\bibitem[Troy et al.(2000)]{troy2000}
         Troy, M., Dekany, R.G., Brack, G., et al. 2000, \procspie, 4007, 31

\bibitem[Valenti \& Fischer(2005)]{valenti2005}
         Valenti, J.A., \& Fischer, D.A. 2005, \apjs, 159, 141

\bibitem[van Leeuwen(2007)]{vanLeeuwen2007}
         van Leeuwen, F. 2007, \aap, 474, 653
 
\bibitem[Vogt et al.(2000)]{vogt2000}
         Vogt, S.S., Marcy, G.W., Butler, R.P., \& Apps, K. 2000, \apj, 536, 902

\bibitem[Wright et al.(2007)]{wright2007}
         Wright, J.T., Marcy, G.W., Fischer, D.A. et al. 2007, \apj, 657, 533

\bibitem[Wu \& Murray(2003)]{wu2003}
         Wu, Y., \& Murray, N. 2003, \apj, 589, 605

\bibitem[Zimmerman et al.(2010)]{zimmerman2010}
         Zimmerman, N., Oppenheimer, B.R., Hinkley, S. et al. \apj, 709, 733

\bibitem[Zimmerman et al.(2011)]{zimmerman2011}
         Zimmerman, N., Brenner, D., Oppenheimer, B. R., Parry, I.R., Hinkley, S., Hunt, S., \& Roberts, R.  2011, \pasp, 123, 746

\bibitem[Zucker \& Mazeh(2002)]{zucker2002} 
         Zucker, S., \& Mazeh, T. 2002, \apjl, 568, L113

\end{thebibliography}
\end{document}